# Spin Hall magnetoresistance in Pt/$Y_3Fe_5O_{12}$ bilayers grown on Si and $Gd_3Ga_5O_{12}$ substrates

Kenta Fukushima[1], Kohei Ueda[1,2,3,a)], Naoki Moriuchi[1], Takanori Kida[4], Masayuki Hagiwara[4], and Jobu Matsuno[1,2,3]

[1]*Department of Physics, Graduate School of Science, Osaka University, Osaka 560-0043, Japan*
[2]*Center for Spintronics Research Network, Graduate School of Engineering Science, Osaka University, Osaka 560-8531, Japan*
[3]*Division of Spintronics Research Network, Institute for Open and Transdisciplinary Research Initiatives, Osaka University, Osaka, 565-0871, Japan*
[4]*Center for Advanced High Magnetic Field Science, Graduate School of Science, Osaka University, Osaka 560-0043, Japan*

**Abstract**

We study spin Hall magnetoresistance (SMR) in Pt/ferrimagnetic insulator $Y_3Fe_5O_{12}$ (YIG) bilayers by focusing on crystallinity, magnetization, and interface roughness by controlling post-annealing temperatures. The SMR in the Pt/YIG grown on Si substrate is comparable to that grown on widely used $Gd_3Ga_5O_{12}$ substrate, indicating that the large SMR can be achieved irrespective to the crystallinity. We deduced the spin mixing conductance from the Pt thickness dependence of the SMR to find the high interface quality of the optimized Pt/YIG grown on Si in terms of spin current. We also clarified that the SMR correlates well with the magnetization, the interface roughness, and carrier density. These findings highlight that optimizing YIG properties is a key to control of magnetization by spin current, leading to the development of low power consumption spintronic device based on the magnetic insulator.

a)e-mail: kueda@phys.sci.osaka-u.ac.jp

Spin Hall effect (SHE)[1] of non-magnet metals (NMs) with strong spin-orbit coupling can convert charge current to spin current. The spin current enables us to manipulate magnetization of a neighboring magnetic layer and hence is essential for developing non-volatile magnetic memory application[2]. The NMs such as Pt[3–6], Ta[7–9], and W[10,11] are effective spin current generators due to their large spin Hall angle ($\theta_{SH}$) which is an efficiency of the charge to spin current conversion via SHE. One of the intriguing phenomena triggered by SHE is spin Hall magnetoresistance (SMR)[12–15] observed in many systems such as NM/ferrimagnet[12,13] and NM/ferromagnet[14,15] bilayers. The SMR measurements have been widely used to extract spin transport parameters in the bilayers such as interfacial spin-mixing conductance ($G^{\uparrow\downarrow}$)[6,13]. Of particular interest in the SMR is use of ferrimagnet insulator $Y_3Fe_5O_{12}$ (YIG). It excludes charge current in the magnetic layer and concomitant shunting effects, making interpretation of the SMR rather simple. With this advantage, a great amount of effort so far has been made in Pt/YIG bilayers[16–30] in order to understand the spin transport at the interface.

The SMR has been demonstrated in epitaxial YIG films grown on $Gd_3Ga_5O_{12}$ (GGG) substrates[12,13,16]. In contrast, the results of spin Seebeck effect (SSE) suggest that thermal spin current is robustly generated from polycrystalline YIG films on the most common substrates Si as well as epitaxial YIG film on GGG substrates[28]. The relationship between spin transport and crystallinity of YIG thus remains an open problem. In order to attack the problem, it is desirable to examine the SMR in the YIG/Si in comparison to YIG/GGG while careful experiments are required since difference of thermal expansion coefficients between YIG and Si causes cracks in the YIG layer[31,32]. Once we obtain large SMR in the YIG/Si, we can fully utilize the advantage of the Si substrate which is compatible with the existing Si technologies in addition to elucidating the SMR mechanism.

In this Letter, we report on the SMR in the Pt/YIG bilayers grown on both GGG and Si substrates by controlling post-annealing temperatures. We found that the SMR strongly depends on the magnetization and interface roughness. With optimized growth conditions, both the YIG/GGG and the YIG/Si films exhibit the same amplitude of the SMR, showing that the SMR is independent of the crystallinity. We further discuss the relation between the film properties and the SMR in both the YIG/Si and YIG/GGG films by estimating the spin mixing conductance from the Pt thickness-dependent-measurement.

We grew YIG films on both the (111)-oriented GGG and the thermally oxidized Si substrates; hereafter we refer to these films as YIG/GGG and YIG/Si, respectively. The YIG films were deposited by RF sputtering at room temperature with an Ar pressure of 1.1 Pa and a sputtering power of 150 W. The as-deposited YIG films were subsequently crystallized by *ex-situ* post-annealing at various temperatures ranging from 700 to 1000°C with 50°C steps in the atmosphere. The crystal structure and the thickness of the YIG/GGG films were confirmed by x-ray diffraction. Figure 1(a) shows $2\theta–\omega$ scan in the YIG film grown on GGG (111) substrate annealed at 750°C, demonstrating clear Laue fringes in addition to GGG (444) peak. Due to the close bulk lattice constants of YIG (1.2376 nm) and GGG (1.2383 nm), diffraction from YIG(444) completely overlaps with that from GGG(444), while an epitaxially grown YIG film with high crystallinity is evidenced by the fringes. By analyzing the fringes, we deduce the out-of-plane lattice constant of the YIG films to be 1.2376 nm, which is in good agreement with the previous report[25]. The high crystallinity of the film is also supported by the full width of half maximum of the rocking curve as narrow as 0.034°. The oscillation also provides the thickness of 54 nm; all the YIG/GGG films maintain the same thickness and crystallinity regardless of the annealing temperatures. Since x-ray diffraction provides very few information on the YIG/Si films due to their lower crystallinity, we characterized the YIG/Si films annealed at 750 and 1000°C by x-ray reflectivity as displayed in Fig. 1(b). From the observed clear fringes, the thickness of the both YIG/Si films were determined to be 55 nm; we confirmed that all the YIG/Si films have the same thickness. The attenuation of intensity in the film annealed at 1000°C is more significant than that in the film annealed at 750°C, manifesting rougher surface in the former film as examined below by atomic force microscope (AFM).

We estimate the saturation magnetization ($M_s$) at 300 K by the superconducting quantum interference device magnetometer. The magnetization curve was measured with sweeping an in-plane magnetic field ($H_{in}$) from +0.4 (+10) to −0.4 kOe (−10 kOe) in YIG/GGG film (YIG/Si films). Figure 1(c) shows a magnetization curve for the YIG/GGG annealed at 750°C, indicating that the $M_s$ of 114 emu/cm$^3$ agrees well with the reported values[16,28,33]. Figure 1(d) shows the corresponding results for the YIG/Si annealed at 750 and 1000°C, which give the $M_s$ = 143 emu/cm$^3$ and 96 emu/cm$^3$, respectively. Surface morphology was estimated by AFM for YIG/GGG annealed at 750°C [Fig. 1(e)] and for YIG/Si annealed at 750 and 1000°C [Fig. 1(f)], respectively. The YIG/GGG film shows a flat surface with root mean square roughness (RMS) of 0.14 nm, comparable to the reported values[24,34]. While the YIG/Si annealed at 750°C has some cracks on the surface with the RMS (0.21 nm) similar to the YIG/GGG, the YIG/Si annealed at 1000°C has a degraded surface with more cracks and even particles, resulting in a poor RMS of 9.2 nm.

We measure the magnetoresistance (MR) of Pt/YIG bilayers deposited on the two different substrates GGG and Si at room temperature. 2-nm-thick Pt Hall bar devices on top of the YIG layer were created via shadow masking the deposition by RF sputtering. Dimensions of the device are 250 µm-width and 625 µm-length, respectively. We measured the longitudinal resistance $R_{xx}$ by rotating a sample with a fixed external magnetic field $H_{ext}$ = 13 kOe and a charge current of 1 mA in three orthogonal planes as shown in Fig. 2(a). The applied $H_{ext}$ is large enough to saturate the $M$ along all coordinate axis. The $R_{xx}$ can be then typically expressed by the following general form[14]:

$$R_{xx} = R_0 - \Delta R_{zy\,(zx)} \sin^2\alpha\,(\beta), \quad (1)$$

where $R_0 = R_{xx}(M // z)$, $\Delta R_{zy(zx)} = R_{xx}(M // z) - R_{xx}(M // y(x))$. We also define $\Delta R_{xy} = R_{xx}(M // x) - R_{xx}(M // y)$, where $\Delta R_{xy}$ corresponds to $\Delta R_{zy} - \Delta R_{zx}$. Here, $\alpha$, $\beta$, and $\gamma$ represent $H_{ext}$ angles in $zy$, $zx$, and $xy$ planes, respectively. The $zy$ scan illustrates the SMR, which is magnetoresistance due to asymmetry between absorption and reflection of spin current generated from the bulk SHE in the NM layer[12–15] [Fig. 2(b)]; this contribution gives higher resistance at $M \parallel z$ and lower resistance at $M \parallel y$, resulting in $\Delta R_{zy} \approx \Delta R_{xy} > 0$ and $\Delta R_{zx} \approx 0$. The $zx$ scan corresponds to anisotropic magnetoresistance (AMR), originating from the enhanced scattering of conduction electrons from the localized $d$- orbitals ($s$-$d$ scattering) in the bulk ferromagnetic metals[35]. This contribution gives $\Delta R_{zx} \approx \Delta R_{xy} > 0$ and $\Delta R_{zy} \approx 0$. Figure 2(c) shows an MR of the YIG/Si annealed at 800°C. We observe the same amplitudes for the $\Delta R_{zy}$ and $\Delta R_{xy}$, and the $\Delta R_{zx} \approx 0$, indicating the sizable SMR and the negligible AMR. Considering the totally insulating nature of the YIG layer, the only possible source of the AMR here is magnetic proximity effect (MPE) at the Pt/YIG interface[18]; the absence of the AMR suggests that the MPE is negligibly small in our case. For further discussion, we focus on the $\Delta R_{zy}$ in order to determine the SMR contribution defined as $\Delta\mathrm{SMR} = \Delta R_{zy}/R_0$ in the right axis of Fig. 2(c). The $\Delta R_{zy}$ and $R_0$ are obtained by the fit using Eq. (1) on the MR curve.

In Figs. 2(e), (g), and (h), we summarize ΔSMR, $M_s$, and RMS as functions of annealing temperature ($T_{ann}$) for the YIG/GGG and YIG/Si films. Note that the reduced $M_s$ value in the YIG/Si film compared to that in the epitaxial YIG/GGG film is possibly from antisite defects which might be enhanced by the epitaxial growth of YIG/GGG in contrast to bulk crystal growth; the magnetization $M_s$ is expected to be reduced when Y ions occupy tetrahedral Fe sites[36,37]. As displayed in Fig. 2(e), the ΔSMR in both the YIG/GGG and the YIG/Si exhibits ~0.15% at a maximum; this is the same as the highest value reported in epitaxial YIG/GGG[16,29,30]. This result suggests that our YIG/Si is as efficient as the epitaxial YIG/GGG in terms of spin injection across the Pt/YIG interface. Based on the SMR theory[12–15], the SMR is proportional to the real part of spin-mixing conductance $G_r^{\uparrow\downarrow}$ under assumption of constant $\theta_{SH}$ in Pt. The $G_r^{\uparrow\downarrow}$ will be discussed later in the paper in the context of spin current through the interface. While the maximum value of the SMR is common in both the YIG/GGG and YIG/Si, we observe a striking contrast in its $T_{ann}$ dependence; ΔSMR shows the maximum values at $T_{ann}$ =750–900°C for the YIG/GGG and $T_{ann}$ =750–800°C for the YIG/Si, indicating a narrow window of optimal $T_{ann}$ for YIG/Si. We attribute this difference to both the $M_s$ and the RMS. While the $M_s$ has the constant value at $T_{ann}$ =750–900°C for both films, it decreases in low and high $T_{ann}$ [Fig. 2(g)]; the reduction is possibly due to the insufficient crystallization[38] at low $T_{ann}$ such as 700°C and the interdiffusion between YIG and substrate at high $T_{ann}$ above 950°C[28], similar to the previous studies of SSE in Pt/magnetic insulators[28,39]. The RMS shown in Fig. 2(h) indicates smooth surface for all the $T_{ann}$ (700–1000°C) in the YIG/GGG films while the RMS becomes rapidly degraded at higher $T_{ann}$ for the YIG/Si films. Since the minimum thickness of the Pt layer in this paper is 1 nm, it is reasonable to define RMS less than 1 nm to be smooth surface which is realized at $T_{ann}$ = 700–800°C for the YIG/Si films. Considering the window of the $M_s$ and the RMS, the films are optimized at $T_{ann}$ = 750–900°C for the YIG/GGG films and at $T_{ann}$ = 750–800°C for the YIG/Si films, which correspond to the regime where the ΔSMR shows a maximum [Fig. 2(e)]. Thus, we evidence that the significantly large SMR can be obtained irrespective to the crystallinity, and that the SMR correlates well with $M_s$, and RMS, by carefully optimizing the YIG film properties. We further comment on relation between the $M_s$ and ΔSMR in the YIG/GGG film. The reduction of the $M_s$ is roughly ~20% from the largest value of 111 emu/cm$^3$ at 750°C to 92 emu/cm$^3$ at 1000°C. This suggests that the drastic decrease of the ΔSMR is related to the reduction of the $M_s$ in the YIG/GGG films as well.

We then perform Hall resistance measurement by sweeping an out-of-plane magnetic field ($H_z$) from +13 kOe to −13 kOe at room temperature. As exemplified in Fig. 2(d), the YIG/Si annealed at 800°C exhibits linear contribution (black dotted lines) from ordinary Hall effect (OHE) in Pt in high $H_z$ regime as well as a small superimposed S-like feature in low $H_z$ regime, that is, an anomalous Hall effect (AHE)-like contribution. We define the contribution as ΔAHE following to Eq. (2).

$$\Delta AHE = \frac{\rho_{xy}(H_z) - \rho_{xy}(-H_z)}{\rho_{xx}^0} \qquad (2)$$

Here, $\rho_{xy}$ and $\rho_{xx}^0$ are the transverse resistivity with $H_z$ and the longitudinal resistivity without $H_z$, respectively. According to the theory of SMR, the imaginary part of the spin mixing conductance $G_i^{\uparrow\downarrow}$ gives rise to spin-Hall induced anomalous Hall effect (SH-AHE) in bilayers including a magnetic insulator[16,22]. In addition, the AHE-like contribution also arises from the MPE at the Pt/magnetic insulator interface, with observation of sizable $\Delta R_{zx}$[18,40]. In our case, we can exclude the MPE contribution as discussed above from the negligible $\Delta R_{zx}$ in Fig. 2(c) and hence the observed AHE is an index of $G_i^{\uparrow\downarrow}$. Figure 2(f) represents ΔAHE versus $T_{ann}$, exhibiting the constant values up to $T_{ann}$ = 900°C and the reduction above $T_{ann}$ = 900°C for the YIG/GGG, while the gradual decrease above 850°C is observed for the YIG/Si; both the SMR and AHE of the YIG/GGG are larger than those of the YIG/Si above 850°C. The difference can be attributed to the RMS [Fig. 2(h)], indicating the role of the interface. This is reasonable since the SMR and SH-AHE represent the interface spin-transport parameters $G_r^{\uparrow\downarrow}$ and $G_i^{\uparrow\downarrow}$, respectively.

For further understanding of the spin current through the Pt/YIG interface, we determine $G_r^{\uparrow\downarrow}$ and $G_i^{\uparrow\downarrow}$ from the Pt thickness ($t$) dependence of the ΔSMR and ΔAHE in Pt($t$ = 1–8nm)/YIG films. Here, we choose the YIG/GGG and the YIG/Si annealed at 750°C as well-optimized films and the YIG/Si annealed at 850°C for comparison. In Fig. 3(a) we find similar $t$-dependent $\Delta$SMR with the maximum value of ~0.15 % at 2 nm in both the YIG/GGG and the YIG/Si annealed at 750°C. The $t$ dependence and the peak agree with the SMR model, which predicts that the SMR takes its maximum at a thickness around $2\lambda$ due to the sufficient spin accumulation[13], where $\lambda$ is the spin diffusion length of NMs layer. Following the model, we fitted these data points by using Eq. (3) with $\theta_{SH} = 0.11$[42].

$$\Delta\text{SMR} = \theta_{SH}^2 \frac{\lambda}{t} \frac{2\lambda G_r^{\uparrow\downarrow}\tanh^2\left(\frac{t}{2\lambda}\right)}{\sigma + 2\lambda G_r^{\uparrow\downarrow}\coth\left(\frac{t}{\lambda}\right)} \quad (3)$$

Here, we use the $t$-dependent electrical conductivity $\sigma$, which is given by $(29\exp^{-t/0.8} + 230)^{-1}$ ($\Omega^{-1}$nm$^{-1}$) obtained from our experimental data with a theoretical model[16,23]. We extracted $\lambda = 1.1$ nm and $G_r^{\uparrow\downarrow} = 6.1\times10^{14}$ $\Omega^{-1}$m$^{-2}$ for both the YIG/GGG and the YIG/Si annealed at 750°C. While this $\lambda$ is comparable to the commonly accepted values for Pt/YIG[16,30,43], the $G_r^{\uparrow\downarrow}$ is larger than any reported values for Pt/magnetic insulators[17,20,44,45]. The largest $G_r^{\uparrow\downarrow}$ demonstrates the significantly large spin injection across the Pt/YIG interface in the well-optimized YIG films. We cannot well fit experimental data in the YIG/Si annealed at 850°C; this may stem from the high resistivity at Pt = 1 nm. As mentioned in the discussion about Fig. 2, the RMS is larger than Pt thickness in case of $T_{ann} = 850$°C and hence we cannot apply Eq. (3). In thicker Pt region, we can roughly estimate $G_r^{\uparrow\downarrow}$ in the YIG/Si with $T_{ann} = 850$°C to be one third of that with $T_{ann} = 750$°C since $G_r^{\uparrow\downarrow}$ is proportional to $\Delta$SMR if $\lambda$ is common. $\Delta$AHE versus $t$ is plotted in Fig. 3(b), exhibiting the similar trend with the $\Delta$SMR versus $t$ [Fig. 3(a)], except for the sign change at 1 nm; the sign change indicates the inversion of $G_i^{\uparrow\downarrow}$ possibly due to some interfacial contributions beyond the well-used formula:

$$\Delta\text{AHE} = \frac{2\lambda^2\theta_{SH}^2}{t} \frac{\sigma G_i^{\uparrow\downarrow}\tanh^2\left(\frac{t}{2\lambda}\right)}{\left[\sigma + 2\lambda G_r^{\uparrow\downarrow}\coth\left(\frac{t}{\lambda}\right)\right]^2} \quad (4)$$

Therefore, we fit data points ranging from 2 to 8 nm using Eq. (4) with $\theta_{SH} = 0.11$ to obtain $\lambda = 0.9$ nm and $G_i^{\uparrow\downarrow} = 5.1\times10^{13}$ ($4.3\times10^{13}$) $\Omega^{-1}$m$^{-2}$ for the YIG/GGG (YIG/Si) at 750°C, and $\lambda = 1.8$ nm and $G_i^{\uparrow\downarrow} = 6.9\times10^{12}$ $\Omega^{-1}$m$^{-2}$ for the YIG/Si at 850°C. We extract $G_i^{\uparrow\downarrow}$ as an index of the interface quality in terms of spin current; the YIG/Si annealed at 750°C is comparable to the epitaxial YIG/GGG while the YIG/Si annealed at 850°C has a degraded interface. This is consistent with the $G_r^{\uparrow\downarrow}$ results, reinforcing the high interface quality of the optimized Pt/YIG grown on Si substrates.

We have discussed in Fig. 2 that the SMR correlates with the saturation magnetization of the magnetic YIG layer and the interface roughness. In order to examine the SMR in terms of the nonmagnetic Pt layer, we plot the Pt carrier density $n$ as a function of the annealing temperature $T_{ann}$ for Pt($t$ = 2 nm)/YIG in Fig. 4(a). The $n$ is estimated from coefficient of the OHE ($c_{OHE}$) shown in Fig. 2(d) by using $n = -1/(ec_{OHE}t)$, where $e$ is the elementary charge. The result indicates that the $n$ is influenced by the $T_{ann}$, similar to the relation between ΔSMR versus $T_{ann}$ in Fig. 2(e). In order to gain more insight into their relation, $n$ versus ΔSMR is displayed in Fig. 4(b), which exhibits that the ΔSMR linearly increases with $n$. Considering that the $n$ is a property of Pt, the linear relation cannot be directly explained by the $T_{ann}$ dependence of the YIG properties in our experimental design. Here, we refer similar results to compare with ours. One is the strong temperature dependence including a sign change of OHE in Pt/YIG bilayer in case of $t$ = 1.5 nm and 2.5 nm[19]; the interfacial electrons from Pt can be affected by the strong exchange interaction within the neighboring YIG layer and hence the density of states at the Fermi level will be possibly modified. Another is the $T_{ann}$ dependence of the SMR in bilayer composed of Pt and a magnetic insulator, $Tm_3Fe_5O_{12}$[41]; additional spin-dependent scattering is expected by the increased Fe impurity concentration in the Pt layer caused by annealing. In addition, it is possible that presence of $PtO_x$ at the interface region also affects the carrier density in higher $T_{ann}$. Considering that the $n$ is low both at 700°C and above 850°C for the YIG/Si, we speculate that the contributions from the Fe impurities and $PtO_x$ are limited. The clear correlation in Fig. 4(b) can be at least related to the change of the electronic structure from the YIG layer; the carrier density in the ultrathin Pt layer is therefore not a bulk property but rather reflecting the interface nature. Since the understanding of the ultrathin Pt on the YIG is still under debate, further investigation is required to clarify the observed correlation of SMR and ordinary Hall effect.

      In conclusion, we have investigated the correlation between the SMR and the film properties of YIG in YIG/Pt bilayers grown on GGG and Si substrates by optimizing annealing temperature. We achieved a significantly large SMR in YIG/Si comparable to widely investigated epitaxial YIG/GGG, suggesting the large spin injection across the YIG/Pt interface with high value of $G_r^{\uparrow\downarrow}$. The SMR has a close relation in the magnetization and the interfacial roughness of YIG, indicating that these properties strongly affect the $G_r^{\uparrow\downarrow}$. The quality of the optimized YIG/Si is confirmed as well by the SH-AHE and the $G_i^{\uparrow\downarrow}$. These findings provide a great advantage for device design toward low power consuming device utilizing charge to spin current conversion.

**Acknowledgement**

      We would like to thank T. Arakawa for technical support. This work was carried out at the Center for Advanced High Magnetic Field Science in Osaka University under the Visiting Researcher's Program of the Institute for Solid State Physics, the University of Tokyo. This work was also supported by the Japan Society for the Promotion of Science (JSPS) KAKENHI (Grant Nos. JP19K15434, JP19H05823, and JP22H04478), JPMJCR1901 (JST-CREST), and Nippon Sheet Glass Foundation for Materials Science and Engineering. We acknowledge stimulating

discussion at the meeting of the Cooperative Research Project of the Research Institute of Electrical Communication, Tohoku University

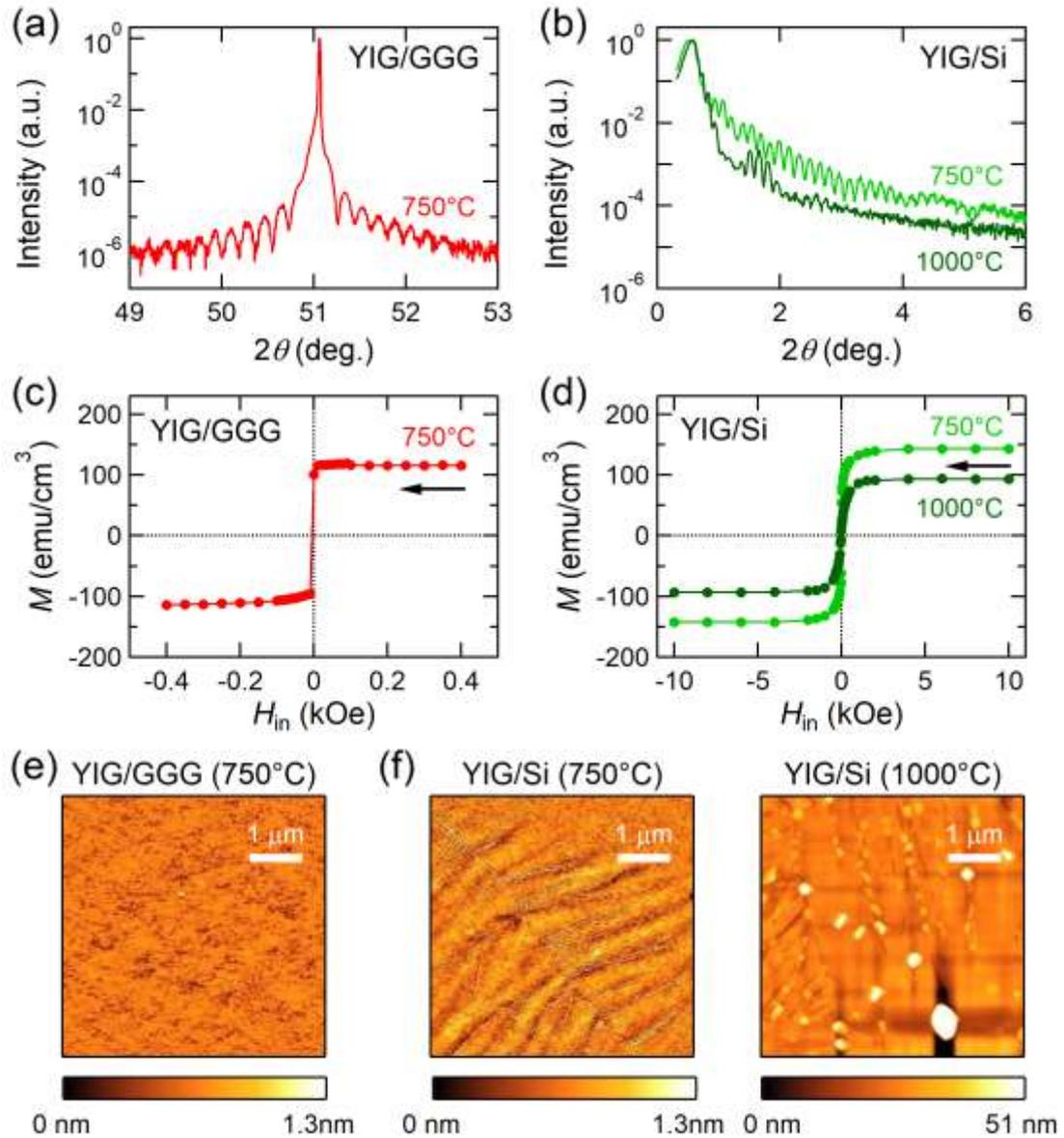

Fig. 1. (a) X-ray diffraction scan (2$\theta$-$\omega$) of the YIG/GGG film annealed at 750°C. (b) X-ray reflectivity of the YIG/Si films annealed at 750 and 1000°C. Magnetization curves measured at 300 K (c) for the YIG/GGG film annealed at 750°C and (d) for the YIG/Si annealed at 750 and 1000°C. The arrows represent the sweeping directions of in-plane magnetic field $H_{in}$. Atomic force microscopy surface images (e) of the YIG/GGG film annealed at 750°C and (f) of the YIG/Si films annealed at 750 and 1000°C.

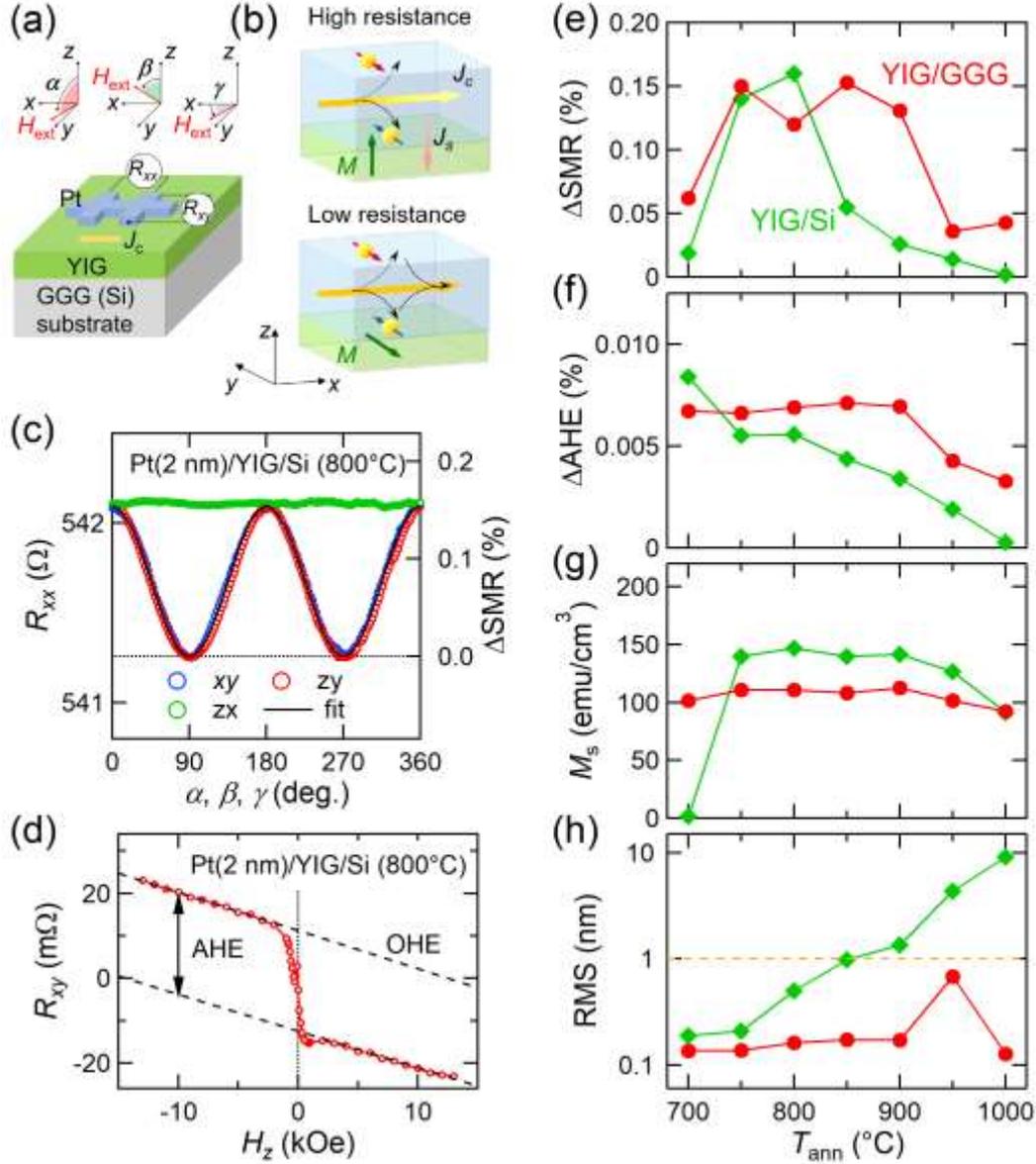

Fig. 2. (a) Schematic illustration of a Hall bar device structure. $\alpha$, $\beta$, and $\gamma$ represent angles of an external magnetic field corresponding to $zy$, $zx$, and $xy$ plane rotations. (b) Schematic illustrations of SMR. $J_c$ and $J_s$ represent charge and spin current, respectively. Upper and lower figures correspond to high resistance state and low resistance state, respectively. (c) Angle dependences of magnetoresistance in the Pt(2 nm)/YIG/Si annealed at 800°C. Black solid line is a fitting result using Eq. (1). (d) An out-of-plane magnetic field ($H_z$) dependence of Hall resistance in Pt(2 nm)/YIG/Si annealed at 800°C. Linear background (dotted lines) gives the contribution of ordinary Hall effect. The contribution of anomalous Hall effect (arrows) is obtained by the two linear backgrounds. (e) ΔSMR, (f) ΔAHE, (g) saturation magnetizations, and (h) RMS roughness as function of annealing temperatures ($T_{ann}$) in both the YIG/GGG and YIG/Si films. The error bars are smaller than the symbol size in (e)–(g).

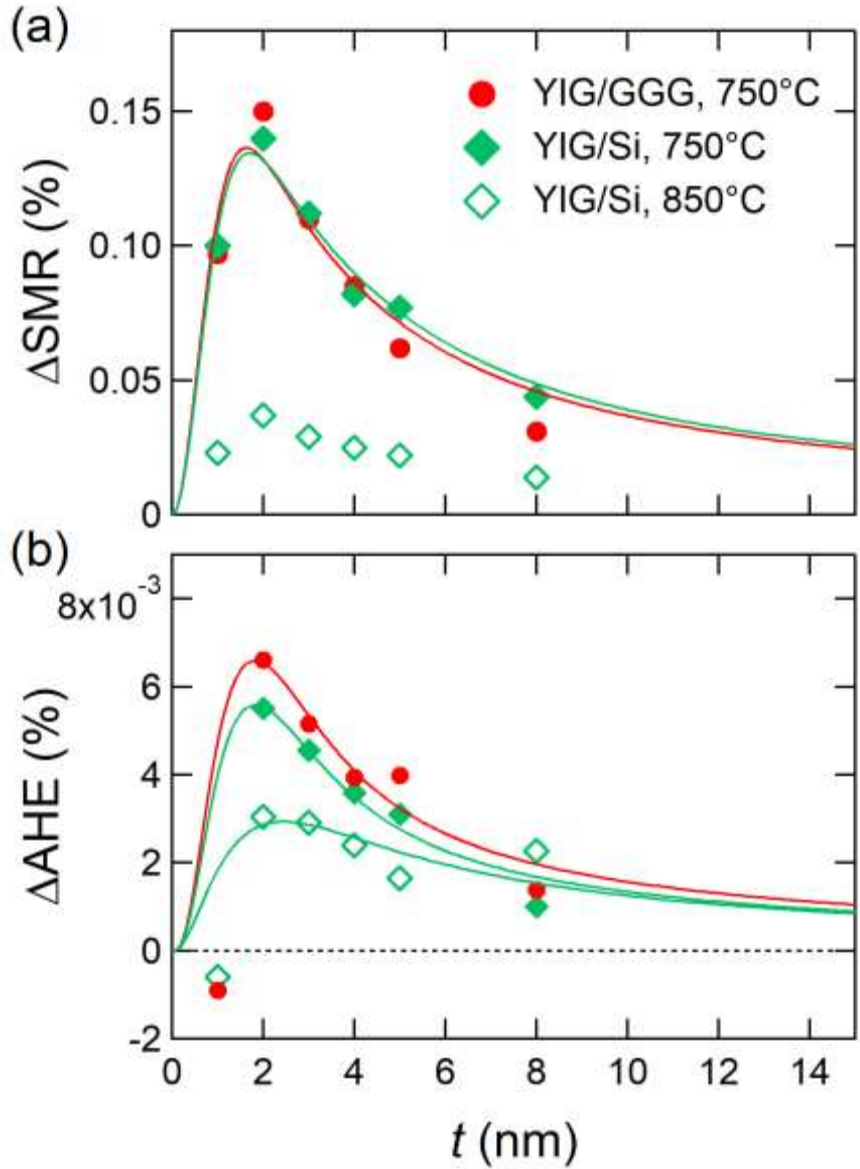

Fig. 3. Pt thickness (*t*) dependence of (a) ΔSMR and (b) ΔAHE in both the YIG/GGG and YIG/Si annealed at 750 °C, and the YIG/Si annealed at 850 °C. Symbols and solid lines represent the experimental data and theoretical curves, respectively. The curves denote the fitting results in each data points from 2 to 8 nm based on Eq. (3) and Eq. (4). The error bars are smaller than the symbol size in (a) and (b).

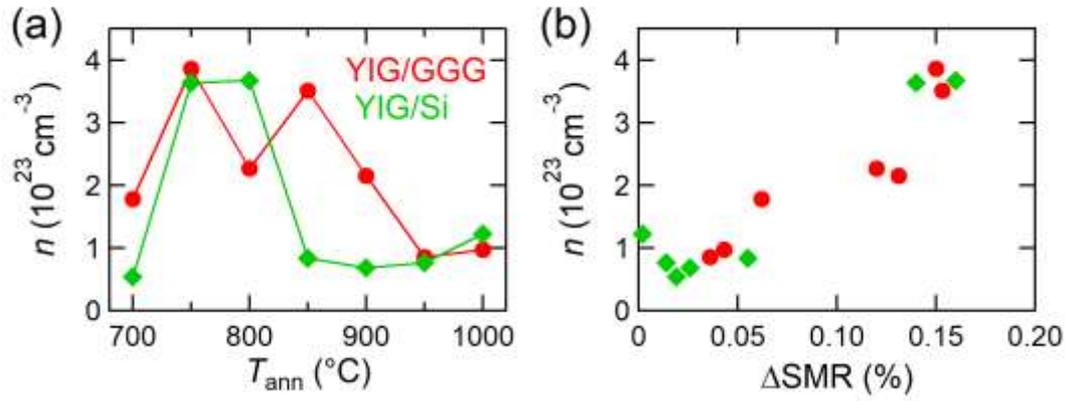

Fig. 4. Carrier density as functions of (a) $T_{ann}$ and (b) $\Delta$SMR in both the YIG/GGG and the YIG/Si films. The error bars are smaller than the symbol size in (a) and (b).

14